\documentclass[twocolumn]{aa}
\usepackage{graphicx}
\usepackage{txfonts}
\begin{document}
\title{Ejecta and progenitor of the low-luminosity Type IIP supernova 2003Z}

\author{Victor P. Utrobin\inst{1,2}, Nikolai N. Chugai\inst{3}, \and
        Andrea Pastorello\inst{4}}

\offprints{V. Utrobin, \email{utrobin@itep.ru}}

\institute{
   Max-Planck-Institut f\"ur Astrophysik,
   Karl-Schwarzschild-Str. 1, D-85741 Garching, Germany
\and
   Institute of Theoretical and Experimental Physics,
   B.~Cheremushkinskaya St. 25, 117218 Moscow, Russia
\and
   Institute of Astronomy of Russian Academy of Sciences,
   Pyatnitskaya St. 48, 109017 Moscow, Russia
\and
   Astrophysics Research Centre, School of Mathematics and Physics,
   Queen's University Belfast, Belfast BT7 1NN, United Kingdom}

\date{Received 6 September 2007 / accepted 28 September 2007}

\abstract{
The origin of low-luminosity Type IIP supernovae is unclear: they have been
   proposed to originate either from massive ($\sim 25~M_{\sun}$) or low-mass
   ($\sim 9~M_{\sun}$) stars.
}{
We wish to determine parameters of the low-luminosity Type IIP supernova 2003Z,
   to estimate a mass-loss rate of the presupernova, and to recover a progenitor
   mass.
}{
We compute the hydrodynamic models of the supernova to describe the light curves
   and the observed expansion velocities.
The wind density of the presupernova is estimated using a thin shell model for
   the interaction with circumstellar matter.
}{
We estimate an ejecta mass of $14.0\pm1.2~M_{\sun}$, an explosion energy of
   $(2.45\pm0.18)\times10^{50}$ erg, a presupernova radius of
   $229\pm39~R_{\sun}$, and a radioactive $^{56}$Ni amount of
   $0.0063\pm0.0006~M_{\sun}$.
The upper limit of the wind density parameter in the presupernova vicinity is
   $10^{13}$ g\,cm$^{-1}$, and the mass lost at the red/yellow supergiant stage
   is $\leq 0.6~M_{\sun}$ assuming the constant mass-loss rate.
The estimated progenitor mass is in the range of $14.4-17.4~M_{\sun}$.
The presupernova of SN~2003Z was probably a yellow supergiant at the time of the
   explosion.
}{
The progenitor mass of SN~2003Z is lower than those of SN~1987A and SN~1999em,
   normal Type IIP supernovae, but higher than the lower limit of stars
   undergoing a core collapse.
We propose an observational test based on the circumstellar interaction to
   discriminate between the massive ($\sim 25~M_{\sun}$) and moderate-mass
   ($\sim 16~M_{\sun}$) scenarios.
}
%
\titlerunning{Parameters of SN~2003Z}
\authorrunning{V. P. Utrobin et al.}
\maketitle

\section{Introduction}
\label{sec:intro}
Type II plateau supernovae (SNe~IIP) with the plateau of $\sim 100$
   days in the light curve are believed to be an outcome of a core collapse of
   the $9-25~M_{\sun}$ stars (e.g., Heger et al. \cite{HFWLH_03}).
This paradigm assuming the Salpeter mass spectrum suggests that about 66\% of
   all SNe~IIP should be produced by progenitors, i.e., stars on the main
   sequence, in the $9-15~M_{\sun}$ range.
At present this general picture remains unconfirmed.
It could be verified via the determination of ejecta masses from the
   hydrodynamic modeling for a sufficiently large sample of SNe~IIP.
Unfortunately, only few SNe~IIP have the well-observed light curves and spectra
   needed to reliably reconstruct the basic SN parameters.
It is not therefore surprising that up to now ejecta mass has been determined
   using the detailed hydrodynamic simulations for only SN~1987A and SN~1999em.
It is noteworthy that, from the point of view of explosion mechanism, SN~1987A
   is a normal SN~IIP; it has the ejecta mass, the explosion energy, and the
   amount of ejected $^{56}$Ni comparable to those of SN~1999em.

Recently an interesting subclass of SNe~IIP, so called low-luminosity SNe~IIP,
   was selected observationally (Pastorello et al. \cite{PZT_04}).
This family is characterized by luminosities, expansion velocities, and
   radioactive $^{56}$Ni masses which are significantly lower than those of
   normal SNe~IIP.
Two views on the origin of low-luminosity SNe~IIP have been proposed.
Turatto et al. (\cite{TMY_98}) have suggested the origin from massive stars,
   $\geq~25~M_{\sun}$, which presumably form black holes and eject a low
   amount of $^{56}$Ni; alternatively, these SNe might originate from
   low-mass stars,  $\sim 9~M_{\sun}$, which are expected to eject a low amount
   of $^{56}$Ni (Chugai \& Utrobin \cite{CU_00}; Kitaura et al. \cite{KJH_06}).

The investigation of the origin of these SNe~IIP began with some confusion.
The point is that SN~1997D, the first low-luminosity SN~IIP, was detected long
   after the explosion and, therefore, was erroneously claimed to possess
   a short ($\sim 50$ days) plateau (Turatto et al. \cite{TMY_98}).
This, in turn, provoked a conclusion that SN~1997D originated from a low-mass
   ($\sim 9~M_{\sun}$) main-sequence star (Chugai \& Utrobin \cite{CU_00}).
The subsequent discovery of several low-luminosity SNe~IIP with a long plateau
   of $\sim 100$ days (Pastorello et al. \cite{PZT_04}) falsified this
   conclusion.
Until now there were no attempts to model hydrodynamically low-luminosity
   SNe~IIP on the basis of new observational data.
The estimate of the ejecta masses of low-luminosity SN~1999br and SN~2003Z was
   made only using a semi-analytical model (Zampieri et al. \cite{ZPT_03};
   Zampieri \cite{Zam_05}).
The semi-analytical model, being a sensible tool for the first order estimates,
   cannot, however, substitute the hydrodynamic simulations.

In this paper we study the well-observed low-luminosity Type IIP SN~2003Z
   (Pastorello \cite{Pas_03}; Knop et al. \cite{KHBD_07}).
Our approach is based on the hydrodynamic modeling of the light curves and
   expansion kinematics.
This paper is organized as follows.
The observational data are presented in Sect.~\ref{sec:obsdat}.
Section~\ref{sec:model} describes the modeling procedure.
Results, specifically, the SN~2003Z parameters and progenitor mass are presented
   in Sect.~\ref{sec:results}.
The implications of the results are discussed in Sect.~\ref{sec:discon}.

A distance to SN~2003Z of 21.68 Mpc is adopted using the Hubble constant
   $H_0=70$ km\,s$^{-1}$\,Mpc$^{-1}$ and a recession velocity of the host
   galaxy NGC 2742 $v_{\mathrm{cor}}=1518$ km\,s$^{-1}$, corrected for the Local
   Group infall to the Virgo cluster and taken from the Lyon Extragalactic Data
   base.
There are no observational signatures of the interstellar absorption in the host
   galaxy (Pastorello \cite{Pas_03}) so a total extinction is taken to be equal
   to the Galactic value $A_{B}=0.167$ (Schlegel et al. \cite{SFD_98}).

\section{Observational data}
\label{sec:obsdat}
SN 2003Z was discovered very young, and a detection limit obtained nine days
   before the discovery allows us to constrain the explosion epoch very well,
   with an uncertainty of only a few days (Boles et al. \cite{BBL_03}).
We adopt an explosion date of SN~2003Z to be JD 2452665.
The spectroscopic and photometric monitoring of SN~2003Z started relatively late,
   about 3 weeks after the discovery, and most of data were collected at
   the 3.5 m Telescopio Nazionale Galileo in La Palma (Spain) and
   the 1.82 m Copernico Telescope in Asiago (Italy).
The plateau phase is quite well sampled (also thanks to the amateur's data),
   and a few additional data were collected during the post-plateau phase.
All data have been reduced using IRAF tasks\footnote {IRAF is distributed by
   the National Optical Astronomy Observatories, which are operated by the
   Association of Universities for Research in Astronomy, Inc., under
   cooperative agreement with the National Science Foundation.}.
In particular, photometric measurements were performed using the point spread
   function fitting technique.
The SN magnitudes were then computed with reference to a sequence of stars in
   the SN vicinity, and the photometric zeropoints of the different nights
   were obtained from observations of several standard fields
   (Landolt \cite{Lan_92}).
The detailed description of the data reduction techniques can be found in
   Pastorello et al. (\cite{PTE_07}).
All observational data of SN~2003Z, first presented in Pastorello
   (\cite{Pas_03}), will be extensively analyzed in a forthcoming paper.

\section{Model overview}
\label{sec:model}
A numerical modeling of SNe~IIP exploits radiation hydrodynamics in the
   one-group approximation (Utrobin \cite{Utr_04}).
When applied to SN~1999em (Utrobin \cite{Utr_07}), the code results in the basic
   SN parameters similar to those recovered in the hydrodynamics with the
   multi-group approach (Baklanov et al. \cite{BBP_05}).
This suggests that the results of our modeling are not hampered by one-group
   approximation of the radiation transfer.

The presupernova (pre-SN) structure is set as a non-evolutionary red supergiant
   (RSG) star in hydrostatic equilibrium.
The adopted composition of the hydrogen envelope is solar.
We admit a mixing between the helium core and the hydrogen envelope similar to
   that in SN~1987A and SN~1999em (Utrobin \cite{Utr_05}, \cite{Utr_07}).
The model explosion is initiated by a supersonic piston applied to the bottom of
   the stellar envelope outside the collapsing $1.4~M_{\sun}$ core.
The time scale for the piston action is $\sim 10^{-3}$~sec.
The explosion energy $E$ is defined as a difference between the total energy
   input and the modulus of the binding energy of the envelope outside the
   collapsing core.
Since in SNe~IIP the radiated energy is small compared to the explosion energy,
   the kinetic energy of the freely expanding ejecta is practically equal to
   the explosion energy.

The strategy of the search for the optimal model is based on the specific
   dependence of the SN light curve and expansion velocities on the model
   parameters (cf. Utrobin \cite{Utr_07}).
The $^{56}$Ni mass is determined by the bolometric luminosity of the
   radioactive tail, while the plateau duration, the plateau luminosity,
   and the velocity at the photosphere depend on the combined effects of the
   ejecta mass, the explosion energy, and the pre-SN radius.
Specifically, the larger ejecta mass results in the longer plateau and the
   lower plateau luminosity; the larger explosion energy leads to the higher
   plateau luminosity and the shorter plateau; the larger pre-SN radius
   results in the higher plateau luminosity and the longer plateau.
The extent of the $^{56}$Ni and helium mixing also affects the light curve,
   particularly the shape of the plateau at the end of the photospheric epoch
   (Utrobin \cite{Utr_07}).
We emphasize the role of the maximal velocity of ejecta in the constraining
   of the model, because the density in the outermost layers is sensitive to
   SN parameters (Utrobin \& Chugai \cite{UC_05}).
Furthermore, the effect of a deceleration of outer layers due to the ejecta/wind
   interaction will be used to estimate the pre-SN wind density as was done
   earlier for SN~1999em (Chugai et al. \cite{CCU_07}).

We rely on the photometric and spectroscopic data reported by Pastorello
   (\cite{Pas_03}).
Although the hydrodynamic model reproduces the $V$, $R$, and $I$ light curves
   satisfactorily, the code with one-group radiation transfer is focused
   on the modeling of the bolometric light curve.
The observed bolometric light curve of SN~2003Z is recovered from $BVRI$
   photometry using the black body approximation for the SN radiation and
   the standard filter responses.
This suggests the determination of the radius of the SN photosphere
   $R_{\mathrm{ph}}$ and the effective temperature $T_{\mathrm{eff}}$
   applying a minimization procedure to the errors between the observed and
   calculated filter responses for the adopted distance and the reddening.
The bolometric luminosity is then calculated as
   $L=4\pi R_{\mathrm{ph}}^2\sigma T_{\mathrm{eff}}^4$, where $\sigma$ is
   the Stephan-Boltzmann constant.
The data on the photospheric velocity evolution are taken from the modeling of
   the Na\,I and Ba\,II lines and from Pastorello (\cite{Pas_03}), while the
   maximal velocity of ejecta, $v_{\mathrm{max}}$, is estimated from the
   H$\alpha$ absorption profile.
The spectrum on day 27.48 (Pastorello \cite{Pas_03}) yields
   $v_{\mathrm{max}}=7500\pm200$ km\,s$^{-1}$.
The same value we derive from the spectrum on day 9 (Knop et al.
   \cite{KHBD_07}).
This indicates a very weak deceleration of the SN outer layers between these
   epochs.

\begin{table}
\caption[]{Physical parameters of hydrodynamic models.}
\label{tab:auxmods}
\centering
\begin{tabular}{c  c  c  c  c  c  }
\hline\hline
\noalign{\smallskip}
 Model & $R_0$ & $M_{env}$ & $E$ & $M_{\mathrm{Ni}}$ & $M_{\mathrm{He}}$  \\
       & ($R_{\sun}$) & ($M_{\sun}$) & ($10^{50}$ erg) & $(10^{-3} M_{\sun})$
        & ($M_{\sun}$)  \\
\noalign{\smallskip}
\hline
\noalign{\smallskip}
Opt & 229 & 14 & 2.45 & 6.28 & 3.9 \\
He8 & 229 & 14 & 2.45 & 6.28 & 8.0 \\
Rms & 195 & 14 & 2.45 & 6.28 & 3.9 \\
Rps & 263 & 14 & 2.45 & 6.28 & 3.9 \\
Mms & 229 & 12 & 2.45 & 6.28 & 3.9 \\
Mps & 229 & 16 & 2.45 & 6.28 & 3.9 \\
Ems & 229 & 14 & 2.10 & 6.28 & 3.9 \\
Eps & 229 & 14 & 2.80 & 6.28 & 3.9 \\
Nms & 229 & 14 & 2.45 & 5.34 & 3.9 \\
Nps & 229 & 14 & 2.45 & 7.22 & 3.9 \\
\noalign{\smallskip}
\hline
\end{tabular}
\end{table}
%

\section{Results}
\label{sec:results}
%
\begin{figure}[b]
   \resizebox{\hsize}{!}{\includegraphics{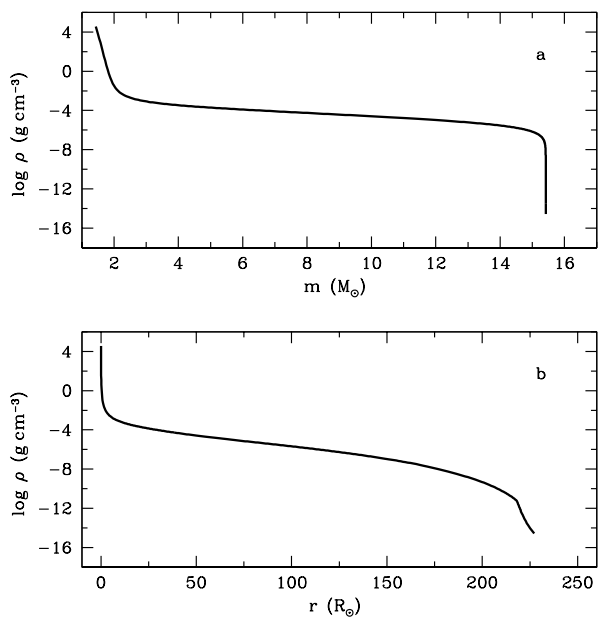}}
   \caption{%
   Density distribution with respect to interior mass (\textbf{a}) and
   radius (\textbf{b}) for the optimal pre-SN model.
   The central core of 1.4 $M_{\sun}$ is omitted.
   }
   \label{fig:denmr}
\end{figure}
\begin{figure}[t]
   \resizebox{\hsize}{!}{\includegraphics[clip, trim=0 0 0 184]{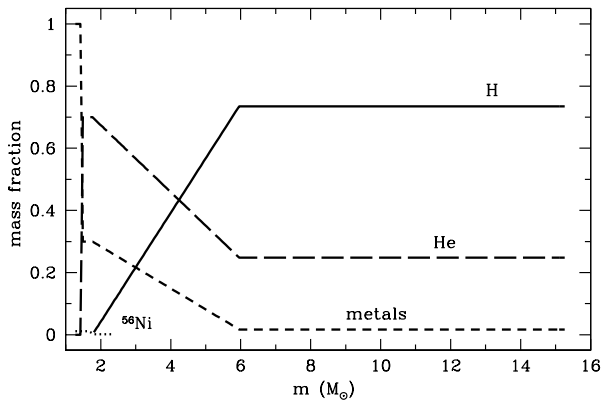}}
   \caption{%
   The mass fraction of hydrogen (\emph{solid line\/}), helium
      (\emph{long dashed line\/}), heavy elements (\emph{short dashed line\/}),
      and radioactive $^{56}$Ni (\emph{dotted line\/}) in the ejecta of
      the optimal model.
   }
   \label{fig:chcom}
\end{figure}
\begin{figure}[t]
   \resizebox{\hsize}{!}{\includegraphics[clip, trim=0 0 0 184]{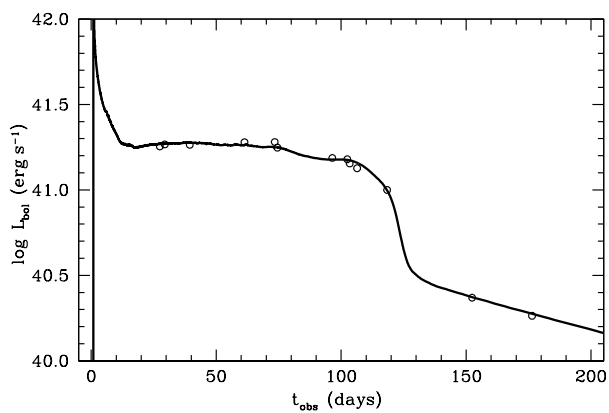}}
   \caption{%
   Comparison of the calculated bolometric light curve of model Opt
      (\emph{solid line\/}) with the bolometric data of SN 2003Z
      evaluated from the photometric observations of Pastorello (\cite{Pas_03})
      (\emph{open circles\/}).
   }
   \label{fig:lmbol}
\end{figure}
%

\subsection{Optimal model}
\label{sec:res-optmod}
An extensive search in the SN parameter space led us to the optimal model of
   SN~2003Z (model Opt in Table~\ref{tab:auxmods}) with the ejecta mass
   $M_{env}=14~M_{\sun}$, the explosion energy $E=2.45\times10^{50}$ erg,
   the $^{56}$Ni mass $M_{\mathrm{Ni}}=0.0063~M_{\sun}$, and the pre-SN radius
   $R_0=229~R_{\sun}$.
The initial density distribution in the model Opt mimics a heterogeneous
   structure of the RSG with the dense metal/helium core and the extended
   hydrogen envelope (Fig.~\ref{fig:denmr}).
The helium core is mixed with the hydrogen envelope so that the helium abundance
   drops linearly along the mass coordinate in the mixing zone
   (Fig.~\ref{fig:chcom}).
In this description the mixing extent is specified by the mass coordinate of
   zero hydrogen abundance ($1.8~M_{\sun}$).
Of note, SN~2003Z similar to previously studied SN~1987A and SN~1999em,
   requires extended mixing between the helium core and hydrogen envelope.

\begin{figure}[t]
   \resizebox{\hsize}{!}{\includegraphics{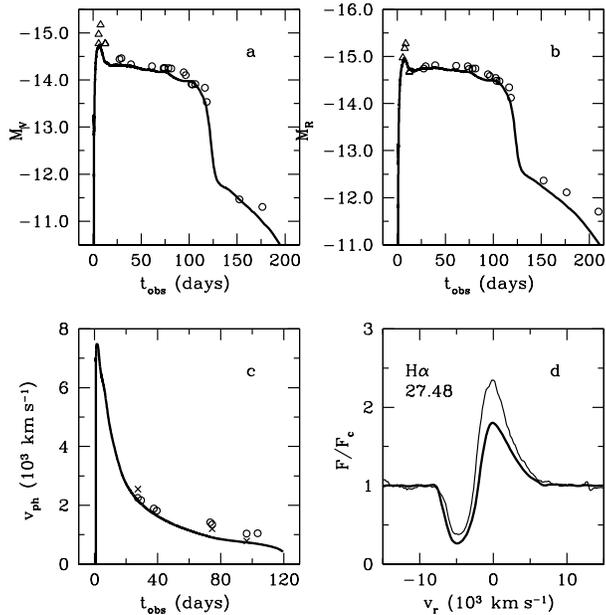}}
   \caption{%
   Optimal hydrodynamic model.
   Panel (\textbf{a}): the calculated $V$ light curve (\emph{solid lines\/})
      compared with the observations of SN 2003Z obtained by Pastorello
      (\cite{Pas_03}) (\emph{open circles\/}) and Gary (\cite{Gary_03})
      (\emph{open triangles\/}).
   Panel (\textbf{b}): the same as panel (a) but for the $R$ light curves.
   Panel (\textbf{c}): calculated photospheric velocity (\emph{solid
      line\/}) is compared with photospheric velocities estimated from
      modeling the Na\,I doublet and Ba\,II 6142 \AA\ profiles (\emph{crosses\/})
      and absorption minima of the Sc\,II 6246 \AA\ line measured by Pastorello
      (\cite{Pas_03}) (\emph{open circles\/}).
   Panel (\textbf{d}): H$\alpha$ profile, computed with the time-dependent
      approach (\emph{thick solid line\/}), overplotted on the observed profile
      on day 27.48, as obtained by Pastorello (\cite{Pas_03}) (\emph{thin
      solid line\/}).
   }
   \label{fig:optmod}
\end{figure}

The bolometric light curve of SN 2003Z is fairly well fitted by the optimal
   model (Fig.~\ref{fig:lmbol}).
Both empirical and calculated $V$ and $R$ light curves (Figs.~\ref{fig:optmod}a
   and \ref{fig:optmod}b) show an initial peak related to the cooling of hot
   outer layers of shocked ejecta.
The amplitude and width of this peak depend on the structure of the outermost
   rarefied layers of the pre-SN envelope.
We did not try to search for the appropriate density structure of the outermost
   layers to reproduce in detail the initial optical peak indicated by
   the amateur observations\footnote{%
   http://www.astrosurf.com/snweb2/2003/03Z\_/03Z\_Meas.htm}.

\begin{figure}[t]
   \resizebox{\hsize}{!}{\includegraphics{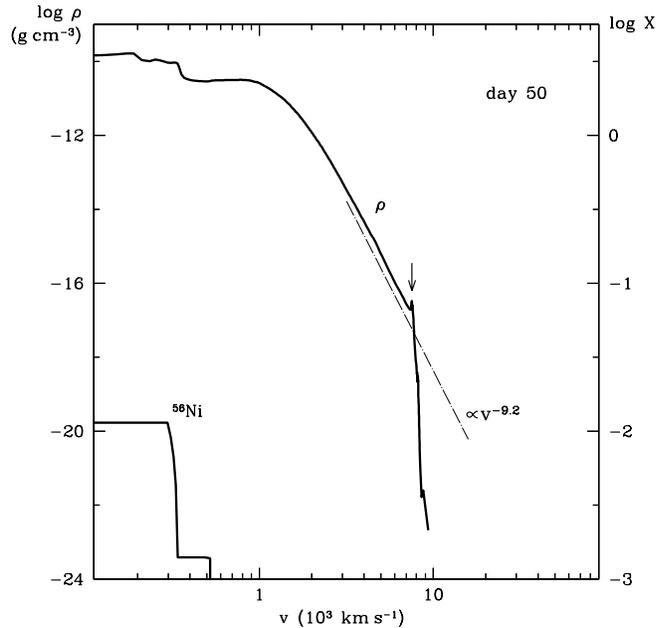}}
   \caption{%
   The density and the $^{56}$Ni mass fraction as a function of the velocity
      for model Opt at $t=50$ days.
   \emph{Dash-dotted line} is the density distribution fit
      $\rho \propto v^{-9.2}$.
   The arrow marks the boundary dense shell at the velocity
      $v_{\mathrm{s}}=7540$ km\,s$^{-1}$.
   }
   \label{fig:denicl}
\end{figure}

The optimal model reproduces satisfactorily the evolution of the photospheric
   velocity (Fig.~\ref{fig:optmod}c).
This evolution is determined primarily by the density distribution in the
   ejecta.
Remarkably, the power low density distribution in the outer layers
   $\rho \propto v^{-n}$ with $n=9.2$ (Fig.~\ref{fig:denicl}) is very close
   to that with the power index $n=9$ preferred by Knop et al.
   (\cite{KHBD_07}) on the basis of the analysis of the early spectra.
The model must also reproduce the maximal velocity
   $v_{\mathrm{max}}=7500\pm200$ km\,s$^{-1}$ derived from the blue
   wing of the H$\alpha$ absorption in the spectrum on day 27.48.
This means that the initial velocity $v_{\mathrm{s}}$ of the boundary shell,
   which forms because of the radiative damping of the shock wave
   (Grassberg et al. \cite{GIN_71}; Chevalier \cite{Che_81}), should
   satisfy the inequality $v_{\mathrm{s}}\geq v_{\mathrm{max}}$.
In the optimal model the boundary dense shell with the mass
   $M_{\mathrm{s}}=1.8\times10^{-4}~M_{\sun}$ is seen in the density
   distribution at the velocity $v_{\mathrm{s}}=7540$ km\,s$^{-1}$
   (Fig.~\ref{fig:denicl}) which meets the above inequality.
It is of note that we failed to find a good hydrodynamic model of SN 2003Z with
   the larger boundary velocity, $v_{\mathrm{s}} > 7540$ km\,s$^{-1}$.
This consideration emphasizes the role of the maximal velocity of ejecta
   in the constraining of the hydrodynamic model.

We calculated the H$\alpha$ line on day 27.48 in the optimal model taking into
   account time-dependent effects of the hydrogen ionization and excitation
   (Utrobin \& Chugai \cite{UC_05}).
The calculated profile reproduces the blue wing of the absorption fairly well
   (Fig.~\ref{fig:optmod}d) although the model is not as good in the emission.
However, the latter is of minor importance for our purpose, the diagnostic of
   the density in outer layers.
Indeed, the absorption depth in the blue wing of the H$\alpha$ line is
   determined primarily by the optical depth which, in the time-dependent
   approach, is a function of density in the outer layers.
The fit of the calculated profile to the observations in the blue absorption
   wing therefore strongly supports the optimal model of SN~2003Z.
The origin of the disagreement in the emission component was studied previously
   for SN~1987A (Utrobin \& Chugai \cite{UC_05}) and it was found to be related
   to our simplified description of the ultraviolet spectrum in the range of
   $h\nu>3.4$ eV.

\begin{figure}[t]
   \resizebox{\hsize}{!}{\includegraphics[clip, trim=0 287 0 0 ]{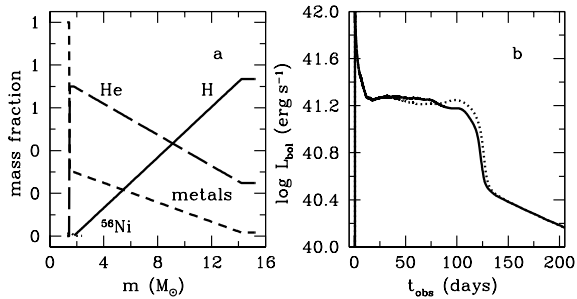}}
   \caption{%
   Dependence on the mass of the helium core.
   Panel~(\textbf{a}): the chemical composition of model He8 that differs
      from model Opt in the mass fraction of hydrogen (\emph{solid line\/}),
      helium (\emph{long dashed line\/}), and heavy elements (\emph{short
      dashed line\/}), the distribution of radioactive $^{56}$Ni
      (\emph{dotted line\/}) being the same.
   Panel~(\textbf{b}): bolometric light curve of model Opt (\emph{thick solid
      line\/}) is compared with that of model He8 (\emph{dotted line\/}).
   }
   \label{fig:hecore}
\end{figure}
The last column in Table~\ref{tab:auxmods} is the total mass of the helium core;
   the value of $3.9~M_{\sun}$ adopted in the optimal model corresponds
   to a $15~M_{\sun}$ progenitor without rotation (Heger et al. \cite{HLW_00}).
The accurate value of the helium core mass is of little significance for the
   optimal model because the results are not sensitive to the helium core in
   a rather broad mass range.
A weak sensitivity to the helium core mass has its dark side, because it
   prevents us from recovering the progenitor mass on the basis of only the
   hydrodynamic modeling.
In this regard it would be of interest to consider the case of a pre-SN with
   a massive helium core of $8~M_{\sun}$ (model He8 in Table~\ref{tab:auxmods})
   which is expected for the $\approx 25~M_{\sun}$ progenitor without rotation
   (Heger et al. \cite{HLW_00}).
The helium core in this model is strongly mixed similar to the optimal
   model (Fig.~\ref{fig:hecore}a).
Although the light curve of model He8 deviates from that of the optimal model
   (Fig.~\ref{fig:hecore}b), the difference is small.
We believe that with a more sophisticated distribution of the mixed helium and
   radioactive $^{56}$Ni one could reach a closer resemblance to the optimal
   model.
This consideration indicates that a massive progenitor ($25-30~M_{\sun}$) for
   SN~2003Z cannot be precluded on the basis of only the hydrodynamic
   modeling.
Note that the ejecta mass for the high-mass scenario should be
   $14~M_{\sun}$, so the star must lose $\Delta M\approx 10~M_{\sun}$ via
   the stellar wind, presumably at the RSG stage during
   $t_{\mathrm{rsg}}\approx4\times10^5$ yr (Heger \cite{Heg_98})
   with a high mass-loss rate
   $\dot{M}=\Delta M/t_{\mathrm{rsg}}\approx2.5\times10^{-5}~M_{\sun}$\,yr$^{-1}$.
This provides us a way to verify the high-mass progenitor using the estimate of the
    wind density around the pre-SN in the combination with the hydrodynamic
    model.

\begin{figure}[t]
   \resizebox{\hsize}{!}{\includegraphics{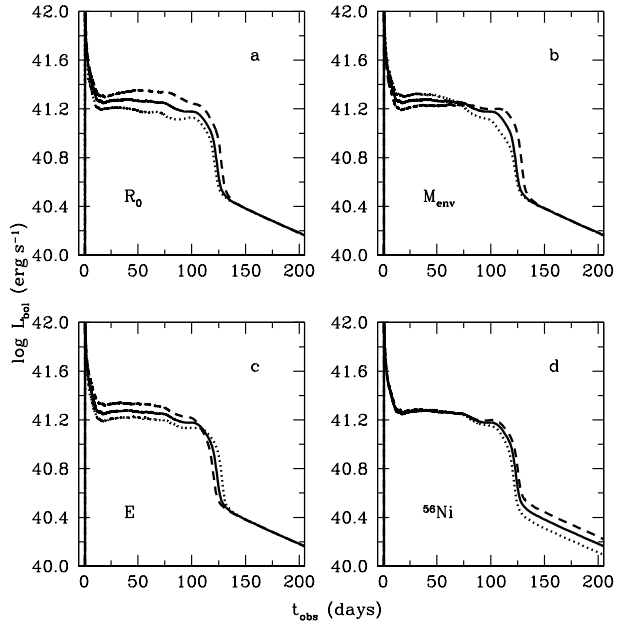}}
   \caption{%
   Dependence of the bolometric light curve of the optimal model
      (\emph{solid line\/}) on the basic parameters:
      (\textbf{a}) the initial radius, model Rms (\emph{dotted line\/})
      and model Rps (\emph{dashed line\/});
      (\textbf{b}) the ejecta mass, model Mms (\emph{dotted line\/})
      and model Mps (\emph{dashed line\/});
      (\textbf{c}) the explosion energy, model Ems (\emph{dotted line\/})
      and model Eps (\emph{dashed line\/});
      (\textbf{d}) the total $^{56}$Ni mass, model Nms
      (\emph{dotted line\/}) and model Nps (\emph{dashed line\/}).
   }
   \label{fig:baspar}
\end{figure}
\begin{table}
\caption[]{Observed parameters of auxiliary hydrodynamic models.}
\label{tab:obspar}
\centering
\begin{tabular}{c c c c c c}
\hline\hline
\noalign{\smallskip}
 Model & $\Delta t$ & $\log L_{bol}^{p}$ & $v_{ph}^{p}$ & $v_\mathrm{s}$ \\
       & (days) & (erg\,s$^{-1}$) & (km\,s$^{-1}$) & (km\,s$^{-1}$) \\
\noalign{\smallskip}
\hline
\noalign{\smallskip}
Opt & 121.51 & 41.2643 & 1117.7 & 7540 \\
Rms & 120.90 & 41.1713 & 1052.2 & 8095 \\
Rps & 124.60 & 41.3429 & 1162.5 & 7497 \\
Mms & 118.58 & 41.2858 & 1132.4 & 8045 \\
Mps & 127.52 & 41.2306 & 1065.4 & 6972 \\
Ems & 126.74 & 41.2047 & 1025.7 & 6963 \\
Eps & 117.32 & 41.3259 & 1224.2 & 8122 \\
Nms & 118.86 & 41.2650 & 1145.6 & 7565 \\
Nps & 123.81 & 41.2614 & 1102.2 & 7573 \\
\noalign{\smallskip}
\hline
\end{tabular}
\end{table}
The confidence in the derived model parameters is determined by the errors in
   the distance, the extinction, the plateau duration, and the velocity at the
   photosphere.
Unfortunately, the errors of the SN~2003Z data are not well defined.
Although the internal accuracy of the photometric data and the velocity
   measurements is better than 5\% (Pastorello \cite{Pas_03}), the luminosity
   error is certainly larger because of uncertainties in the distance and the
   extinction, while the photospheric velocity measured from spectral lines and
   the calculated photospheric velocity may refer to physically different layers.
We therefore adopt rather arbitrarily, the following relative errors: 10\% in the
   bolometric luminosity, 5\% in the photospheric velocity, and 5\% in the
   plateau duration.
Note that the uncertainty of the $^{56}$Ni mass is determined by the error in
   the bolometric luminosity.
At first glance the error in the luminosity is underestimated, since the error
   in the distance could be as large as 10\% (or 20\% in luminosity) given
   possible uncertainties in the Hubble constant and in the corrected recession
   velocity of NGC 2742.
On the other hand, our extensive modeling shows that in order to reproduce both
   the observed maximal ejecta velocity and the light curve, one needs to adopt
   the maximal distance to NGC 2742 (21.68 Mpc).
No much room, therefore, is left for the distance variation.
This justifies our choice for the luminosity error.

Using the auxiliary models (Tables~\ref{tab:auxmods} and \ref{tab:obspar},
   Fig.~\ref{fig:baspar}), we are able to translate the adopted errors of the
   observational parameters into the uncertainties in the pre-SN radius of
   $\pm39~R_{\sun}$, the ejecta mass of $\pm1.2~M_{\sun}$, the explosion energy
   of $\pm0.18\times10^{50}$ erg, and the $^{56}$Ni mass of
   $\pm0.0006~M_{\sun}$.
The ejecta mass thus turns out to be in the range $12.8-15.2~M_{\sun}$.
Amazingly, using the semi-analytical model, Zampieri (\cite{Zam_05}) found that
   the ejecta of SN~2003Z lies in the range $13-19~M_{\sun}$ which is very close
   to our result.
With the adopted errors, the pre-SN mass (ejecta plus neutron star) falls into
   the $14.2-16.6~M_{\sun}$ range.

\subsection{Progenitor}
\label{sec:res-progen}
%
\begin{figure}[t]
   \resizebox{\hsize}{!}{\includegraphics[clip, trim=0 0 0 184]{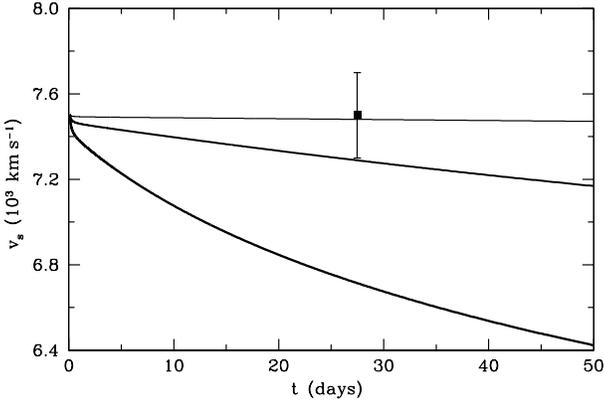}}
   \caption{%
   Evolution of the maximal velocity of the ejecta in the optimal model
      for three values of dimensionless wind density parameter $\omega$:
      1 (\emph{thick line\/}), 0.15 (\emph{medium line\/}), and 0.01
      (\emph{thin line\/}).
   The observed maximal velocity with the corresponding uncertainty is
      shown on day 27.48.
   }
   \label{fig:vmax}
\end{figure}
The small difference between the computed boundary shell velocity of $7540$
   km\,s$^{-1}$ and the observed maximal velocity on day 27.48 of
   $7500\pm200$ km\,s$^{-1}$ places a large constraint on the pre-SN
   wind density: the absence of the pronounced ejecta deceleration during
   the first 27 days implies that the wind should not be very dense.
To find an upper limit of the wind density parameter $w=\dot{M}/u$
   (where $u$ is the wind velocity), we compute the deceleration of
   the outer layers of the model ejecta assuming the thin shell approximation
   (Chugai et al. \cite{CCU_07}).
We introduce for convenience the dimensionless parameter $\omega=w/w_1$ where
   $w_1=6.3\times10^{13}$ g\,cm$^{-1}$; the latter corresponding, for example, to the
   combination of $\dot{M}=10^{-6}~M_{\sun}$\,yr$^{-1}$ and $u=10$ km\,s$^{-1}$.
The calculated evolution of the maximal velocity of the SN~2003Z ejecta for
   the optimal model is shown in Fig.~\ref{fig:vmax} for three cases:
   $\omega=1$, 0.15, and 0.01.
The case $\omega=1$ predicts a too strong deceleration and should be discarded.
The wind with $\omega=0.15$ is barely consistent with the observed maximal
   velocity within the error.
We thus conclude that the density parameter for the pre-SN wind of SN~2003Z is
   $\omega \leq 0.15$ or $w \leq 10^{13}$ g\,cm$^{-1}$ in CGS units.

Another look at the wind density issue could be provided by the analysis of
   the contribution of the circumstellar interaction to the late time luminosity
   of SN 2003Z.
Unfortunately, the photometry for this SN is limited by the age of $\approx 200$
   days at which only a high upper limit for the density of the circumstellar
   matter can be derived.
We can consider, however, SN~1997D, another low-luminosity SN~IIP, which does
   not show signatures of the circumstellar interaction in the light curve till
   day 400 (Benetti et al. \cite{BTB_01}).
This suggests the absence of a dense circumstellar medium around this SN.
We can use the circumstellar interaction model (Chugai et al. \cite{CCU_07})
   to quantify this statement.
Given the similarity of SN~1997D and SN~2003Z we use the ejecta mass and kinetic
   energy of SN~2003Z to estimate the upper limit of the wind density parameter
   $w \sim 6\times10^{14}$ g\,cm$^{-1}$ for SN~1997D.
It is natural to admit that a similar wind is characteristic of the SN~2003Z
   environment as well.
This upper limit, however, is of little interest in the context of SN~2003Z
   because it is 60 times higher than that obtained above from the early time
   ejecta deceleration.

The wind density parameter of SN~2003Z estimated from the ejecta deceleration is
   at least seven times lower than that for the normal Type IIP SN~1999em
   (Chugai et al. \cite{CCU_07}).
Assuming that the wind velocities of both pre-SNe are comparable we conclude
   that the mass-loss rate in the case of SN~2003Z is at least seven times lower
   than that of SN~1999em.
Given that the mass-loss rate increases with the progenitor mass
   (Neuwenhujzen \& de Jager \cite{NJ_90}) the difference in the wind density
   implies that the progenitor of SN~2003Z is less massive compared to
   that of SN~1999em.

In order to convert the pre-SN mass into the progenitor mass, requires
   knowledge of the mass lost via the stellar wind on the main sequence and at
   the RSG stage.
The mass lost on the main sequence for the $\sim 15~M_{\sun}$ star is
   $\sim 0.2~M_{\sun}$ (cf. Heger et al. \cite{HLW_00}), while at the RSG
   stage the lost mass is $\Delta M=t_{\mathrm{rsg}}wu$,
   where $t_{\mathrm{rsg}}\sim10^6$ yr is the duration of the RSG stage
   for the $\sim 15~M_{\sun}$ star (Heger \cite{Heg_98}).
The wind velocity is a poorly known parameter.
For the RSG stars the velocity is in the range of $10-40$ km\,s$^{-1}$
   (Cherchneff \& Tielens \cite{CT_94}).
However, the pre-SN of SN~2003Z with the radius of $229~R_{\sun}$ seems to be
   closer to a yellow supergiant (YSG).
Indeed, for the $15~M_{\sun}$ star luminosity of $L\approx6\times10^{4}~L_{\sun}$
   (Heger et al. \cite{HLW_00}), the effective temperature of the pre-SN of
   SN~2003Z would be $T_{\mathrm{eff}}\approx6000$~K, i.e., typical for a YSG.
The wind velocities estimated from the YSG observations fall into the range of
   $30-40$ km\,s$^{-1}$ (Lobel et al. \cite{LDS_03};
   Robberto et al. \cite{RFN_93}).
With the maximal velocity $u=40$ km\,s$^{-1}$ the mass-loss rate immediately
   before the explosion is $\dot{M}\leq 0.6\times10^{-6}~M_{\sun}$ yr$^{-1}$.
We assume that the mass-loss rate during the whole RSG/YSG stage is constant
   on average in agreement with the general wisdom that the value of $\dot{M}$ is
   mainly determined by the stellar mass and luminosity (Neuwenhujzen \&
   de Jager \cite{NJ_90}).
The mass lost at the RSG/YSG stage is then $\leq 0.6~M_{\sun}$.
Adopting the mass of $0.2~M_{\sun}$ to be lost on the main sequence, we conclude
   that the progenitor lost $0.2-0.8~M_{\sun}$ owing to the wind.
Combined with the pre-SN mass of $15.4\pm1.2~M_{\sun}$ this implies the
   progenitor mass of SN~2003Z to be in the range of $14.4-17.4~M_{\sun}$
   with the average value of $\approx 16~M_{\sun}$.

\section{Discussion and conclusions}
\label{sec:discon}
The goal of this study was to recover the basic parameters of SN 2003Z and
   to get an idea about progenitor masses of low-luminosity SNe~IIP.
We estimated the ejecta mass to be $14.0\pm1.2~M_{\sun}$, the explosion energy
   $(2.45\pm0.18)\times10^{50}$ erg, the pre-SN radius $229\pm39~R_{\sun}$,
   and the $^{56}$Ni mass $0.0063\pm0.0006~M_{\sun}$.
Using the ejecta/wind interaction model, we found the upper limit of the
   density parameter of the pre-SN wind and, assuming the constant mass-loss
   rate at the RSG/YSG stage, constrained the progenitor mass by the range of
   $14.4-17.4~M_{\sun}$.
The estimate of the wind density thus allows us to avoid the
   uncertainty in the progenitor mass stemming from the weak sensitivity of
   the hydrodynamic model to the helium core mass.

\begin{figure}[t]
   \resizebox{\hsize}{!}{\includegraphics{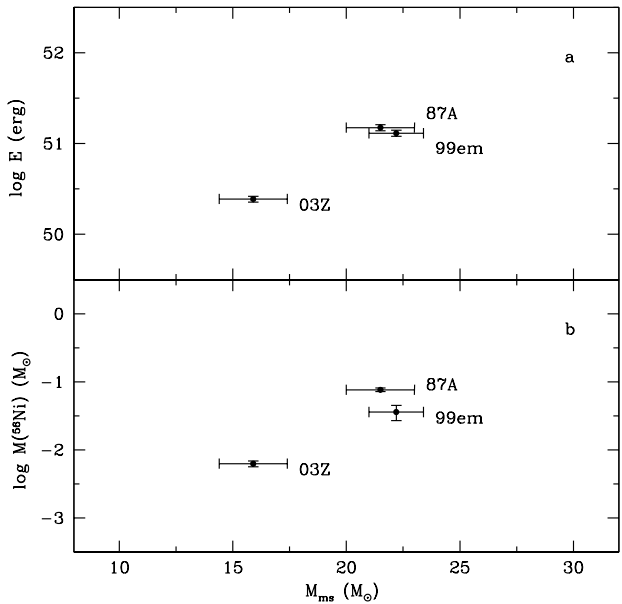}}
   \caption{%
   Explosion energy (\textbf{a}) and $^{56}$Ni mass (\textbf{b}) versus
      progenitor mass for three core-collapse SNe.
   }
   \label{fig:nienms}
\end{figure}

The only normal SN~IIP, studied in a similar way to SN~2003Z, is SN~1999em.
Its pre-SN radius is $\approx 500~R_{\sun}$, the ejecta mass is
   $\approx 19~M_{\sun}$, the explosion energy is
   $\approx 1.3\times10^{51}$ erg, and the $^{56}$Ni mass is
   $\approx 0.036~M_{\sun}$ (Utrobin \cite{Utr_07}), while the progenitor mass
   is $\approx 22~M_{\sun}$ (Chugai et al. \cite{CCU_07}).
A comparison of SN~2003Z with SN~1999em taken together with the fact that the
   low-luminosity SNe~IIP are very similar indicates that this variety of
   SNe~IIP originates from less massive progenitors than normal SNe~IIP.
Parameters of three SNe~IIP --- SN~1987A (Utrobin \cite{Utr_05}), SN~1999em
   (Utrobin \cite{Utr_07}; Chugai et al. \cite{CCU_07}), and SN~2003Z
   (Table \ref{tab:sumtab}) --- determined on the basis
   of the similar hydrodynamic modeling, suggest a picture in which ordinary
   SNe~IIP originate from massive progenitors around $20~M_{\sun}$,
   while low-luminosity SNe~IIP originate from less massive progenitors around
   $\sim 16~M_{\sun}$, close to the average mass of the $9-25~M_{\sun}$ range
   traditionally associated with SNe~IIP.
If our result for SN~2003Z is correct, then the explosion energy and the
   amount of ejected $^{56}$Ni should significantly increase for the progenitors
   between $\sim 16~M_{\sun}$ and $\sim 20~M_{\sun}$ (Fig.~\ref{fig:nienms}).
Interestingly, the empirical correlation between the explosion energy and
   the $^{56}$Ni mass for normal SNe~IIP has been demonstrated by
   Nadyozhin (\cite{Nad_03}).
It should be emphasized that these correlations, valid for SNe~IIP, may not be
   applicable to other SNe~II.
At least SN~1994W (Type IIn event) was found to have a low amount of ejected
   $^{56}$Ni, $\sim 0.015~M_{\sun}$, but a normal explosion energy,
   $\approx 1.3\times10^{51}$ erg (Chugai et al. \cite{CBC_04}).

\begin{table}[t]
\caption[]{Hydrodynamic models for SN 1987A, SN 1999em, and SN~2003Z.}
\label{tab:sumtab}
\centering
\begin{tabular}{@{ } c @{ } c @{ } c @{ } c @{ } c @{ } c @{ } c @{ } c @{ }}
\hline\hline
\noalign{\smallskip}
 SN & $R_0$ & $M_{env}$ & $E$ & $M_{\mathrm{Ni}}$ & $v_{\mathrm{Ni}}^{max}$
       & $v_{\mathrm{H}}^{min}$ & $M_\mathrm{ms}$ \\
       & $(R_{\sun})$ & $(M_{\sun})$ & ($10^{51}$ erg) & $(10^{-2} M_{\sun})$
       & (km\,s$^{-1}$) & (km\,s$^{-1}$) & $(M_{\sun})$ \\
\noalign{\smallskip}
\hline
\noalign{\smallskip}
 87A &  35 & 18 & 1.5   & 7.65 & 3000 & 600 & 21.5 \\
99em & 500 & 19 & 1.3   & 3.60 &  660 & 700 & 22.2 \\
 03Z & 229 & 14 & 0.245 & 0.63 &  535 & 360 & 15.9 \\
\noalign{\smallskip}
\hline
\end{tabular}
\end{table}
Two alternative conjectures about the origin of low-luminosity SNe~IIP have been
   proposed: (1) progenitors are massive stars, $\geq 25~M_{\sun}$
   (Turatto et al. \cite{TMY_98}); (2) these SNe originate from low-mass
   stars, $\sim 9~M_{\sun}$ (Chugai \& Utrobin \cite{CU_00};
   Kitaura et al. \cite{KJH_06}).
The present estimate of the progenitor mass of SN~2003Z, $14.4-17.4~M_{\sun}$,
   is in the disparity with both the high-mass and low-mass scenarios.

Our estimate of the progenitor mass is essentially based on the assumption
   that the mass-loss rate at the RSG/YSG stage was constant.
In fact, the derived wind density refers to the close vicinity of SN~2003Z
   $r<R_{\mathrm w}=vt\sim 10^{16}$ cm (where $v\approx7\times10^8$ cm s$^{-1}$
   is the SN expansion velocity in the outer layers and $t\sim10^7$ sec is the
   characteristic age of the SN).
This linear scale with the wind velocity $u>10$ km\,s$^{-1}$ corresponds to
   the wind history during the latest $R_{\mathrm w}/u <300$ yr before
   the explosion.
Most of the RSG/YSG stage, i.e. $4\times10^5-10^6$ yr, is thus hidden from
   our sight.
One might suggest that the mass-loss rate was substantially more vigorous
   in the past than immediately before the SN explosion.
If this questionable possibility were the case, the $25~M_{\sun}$ progenitor
   would lose $\sim 10~M_{\sun}$, and the high-mass scenario for SN~2003Z would
   be saved.
Of note, the essentially higher wind density of pre-SN in this case suggests
   that the dense wind at the large radii could be revealed via radio and X-ray
   observations at the large age, $t\geq10$ yr.

A remarkable property of all three SNe is strong mixing between the helium core
   and the hydrogen envelope indicated by a low minimal velocity of hydrogen
   matter (Table \ref{tab:sumtab}).
Generally, the model with the unmixed helium core shows a bump in the light
   curve at the end of the plateau (Utrobin \cite{Utr_07}); the absence of
   the bump in the available light curves of SNe~IIP, in addition to the
   narrow-topped H$\alpha$ emission at the nebular phase, indicates that
   substantial mixing between the helium core and the hydrogen envelope is
   a universal phenomenon for SNe~IIP.
This mixing could be induced by the convection during the growth of helium core
   (Barkat \& Wheeler \cite{BW_88}) or/and during the SN explosion (Kifonidis
   et al. \cite{KPSJM_06}).

It is noteworthy that the radii of all three pre-SNe~IIP (Table \ref{tab:sumtab}) are
   notably lower than a radius of typical massive RSG, e.g., the radius of
   $\approx 800~R_{\sun}$ for $\alpha$ Ori (Harper et al. \cite{HBL_01}).
We have already shown that the SN~2003Z pre-SN was the YSG rather than the RSG.
For SN~1999em the $22~M_{\sun}$ pre-SN with the luminosity of
   $\approx 10^5~L_{\sun}$ is characterized by the effective temperature of
   $\sim 4700$~K, which is intermediate between the RSG and YSG values.
Interestingly, the pre-SN of SN~2004et, a normal SN~IIP, was probably a YSG
   as well (Li et al. \cite{LVFC_05}).
These cases suggest that not only the pre-SN~1987A, but a possibly
   significant fraction of pre-SNe~IIP, experience an excursion on the
   Hertzsprung-Russell (HR) diagram from the RSG towards the YSG before
   the explosion.
Note, some pre-SNe~IIP, when becoming the YSG, could get into the instability
   strip and manifest themselves before the explosion as long period
   ($P\sim50-100$ days) Cepheids.
If this is the case, the recovery of the Cepheid period of a pre-SN on the
   basis of a set of pre-explosion observations could provide us with an
   additional tool for the mass determination from the pulsation period.

The recovered progenitor mass of SN~2003Z raises a crucial
   question: what happens to the $9-15~M_{\sun}$ stars which make up
   $\approx66$\% of all the stars from the $9-25~M_{\sun}$ mass range.
Two conceivable suggestions are: (A) all the $9-15~M_{\sun}$ stars
   explode as low-luminosity, or even fainter, SNe~IIP;
   (B) low-luminosity SNe~IIP originate only from a narrow range of
   progenitors around $\sim 16~M_{\sun}$, while the rest of the $9-15~M_{\sun}$
   stars produce different varieties of SNe~II.
A large sample of SNe~II which is well observed and studied is certainly needed to
   distinguish between the options A and B.
It should be emphasized that the initial peak, the end of the plateau, and the
   radioactive tail are of crucial importance for the recovery of reliable
   SN~II parameters.

An alternative approach to determine the pre-SN mass exploits archival
   pre-explosion images of host galaxies and stellar evolution tracks
   on the HR diagram.
Interestingly, using the archival HST images Van Dyk et al. (\cite{VLF_03}) and
   Maund \& Smartt (\cite{MS_05}) derived an upper limit of $\sim 15~M_{\sun}$
   and $\sim 12~M_{\sun}$, respectively, for the progenitor of the
   low-luminosity SN~1999br.
Given a similarity of known low-luminosity SNe~IIP these estimates make
   the high-mass scenario unlikely.

It should be emphasized that a test of the mass determination method based on
   the pre-explosion images is needed; for instance, the pre-SN light could be
   partially absorbed in a dusty circumstellar envelope.
In this regard it would be of top priority to determine independently the
   progenitor mass both from the pre-discovery images and from the hydrodynamic
   simulations.
Among SNe~IIP, except for SN~1987A, only SN~2004et and SN~2005cs were detected
   in the pre-discovery images (Li et al. \cite{LVFC_05};
   Maund et al. \cite{MSD_05}; Li et al. \cite{LVF_06}) and have the light
   curves and spectra appropriate for the hydrodynamic modeling as well
   (Sahu et al. \cite{SASM_06}; Misra et al. \cite{MPC_07};
   Pastorello et al. \cite{PST_06}; Tsvetkov et al. \cite{TVS_06}).
These two challenging cases require a detailed study.

\begin{acknowledgements}
VU thanks Wolfgang Hillebrandt for the excellent opportunity to work
   at the MPA.
We thank the referee David Branch for helpful comments.
This work was supported in part by the Russian Foundation for Fundamental
   Research (05-02-17480 and 08-02-00761).
\end{acknowledgements}


\end{document}